\def\eqnum#1{\eqno (#1)}
\def\fnote#1{\footnote}

\def\square{\hbox{\vrule\vbox{\hrule\phantom{o}\hrule}\vrule}}
\documentstyle[11pt]{article}
\begin{document}
\bigskip
\centerline{{\Large ROTATING DUST SOLUTIONS OF EINSTEIN'S EQUATIONS}}
\medskip
\centerline{{\Large WITH 3-DIMENSIONAL SYMMETRY GROUPS}}
\medskip
\centerline{{\Large PART 1: TWO KILLING FIELDS SPANNED ON $u^{\alpha}$ AND $w^{\alpha }$}}
\bigskip
\par
\centerline{Andrzej Krasi\'nski
}
\centerline{N. Copernicus Astronomical Center and College of Science}
\centerline {Polish Academy of Sciences, Bartycka 18, 00 716 Warszawa, Poland}
\par
\centerline{email: akr@alfa.camk.edu.pl}
\bigskip
{\bf Abstract.} For a rotating dust with a  3-dimensional  symmetry
group all possible metric forms can  be  classified  and,  within
each class, explicitly written out. This is made possible by  the
formalism of Pleba\'nski based  on  the  Darboux  theorem.  In  the
resulting coordinates, the Killing vector fields (if  any  exist)
assume a special form. Each Killing vector field  may  be  either
spanned on the  fields  of  velocity  and  rotation  or  linearly
independent of them. By considering all such cases one arrives at
the classification. With respect to the structures of the groups,
this is just the Bianchi classification, but  with  all  possible
orientations of the orbits taken into  account.  In  this  paper,
which is part 1 of a 3-part series, all solutions are  considered
for  which  two  Killing  fields  are  spanned  on  velocity  and
rotation. The solutions of Lanczos and G\"{o}del are identified as special cases, and their new invariant definitions are provided. In addition, a new invariant definition is given  of  the
Ozsvath class III solution.
\par
\medskip
{\bf 1. Introduction and summary.}
\par
The theorem of Darboux presented in  sec.  2  allows  one  to
introduce invariantly defined coordinates in which  the  velocity
field  of  a  fluid  (not  assumed  to  be  perfect)  acquires  a
"canonical" form. In this paper it is further  assumed  that  the
fluid moves with zero acceleration and  nonzero  rotation.  These
assumptions result in a simplification of the metric  tensor  and
in limitations imposed on the  Killing  vectors,  if  any  exist.
Within this special class  of  coordinates,  any  single  Killing
field may also be reduced to a "canonical" form, a different  one
in the case when it is spanned on the vector fields  of  velocity
$u^{\alpha }$ and rotation $w^{\alpha }$, and a  different  one  when  it  is  linearly
independent of $u^{\alpha }$ and $w^{\alpha }$. This gives rise to a classification  of
possible symmetries in rotating matter.
\par
When there exist three linearly independent  Killing  fields,
the classification described  above  gives  rise  to  a  complete
classification of all possible metric forms. With respect to  the
algebras of  the  symmetry  groups,  this  is  just  the  Bianchi
classification, but with all orientations of the  orbits
in the spacetime taken into account. As a by-product, all the Bianchi types that
are compatible with rotating dust, with any  orientation  of  the
orbit, are identified.
\par
In every case that emerges, the commutation relations of  the
algebra have been solved, resulting in explicit formulae for  the
Killing fields, and then the Killing equations have been  solved,
resulting in the formulae for the metric tensors compatible  with
the symmetry groups considered. The degree of success  in  solving
the Einstein equations varied very strongly from case to case. In
most cases, no headway was  made.  In  some  cases  the  Einstein
equations have been integrated either to  an  autonomous  set  of
first order equations  or  to  a  single  nonlinear  differential
equation of second or third order. In a few cases solutions known
earlier were identified in the present scheme and new invariant
definitions for some of them 
were provided (those by Lanczos [1] and  G\"odel [2] will be mentioned in this paper). In just one case a new solution was found.
\par
Since the number of cases is rather large, the  results  will
be presented in three papers. The present paper 1 deals with  the
simplest situation when two of the Killing fields are spanned  on
velocity and rotation (the case of all three Killing fields being
spanned on $u$ and $w$ is trivial - see sec. 6).
\par
The  Darboux  theorem  was  first  applied  as  a  tool   for
investigating the equations of motion and the Einstein  equations
by Pleba\'nski [3]. He showed that if a perfect fluid  is  rotating
and isentropic while the particle number  is  conserved,  then  a
similar consideration to the  one  presented  here  applies.  The
approach of Pleba\'nski was used by this author [4 - 8] to  find  a
large collection of stationary, cylindrically symmetric solutions
of Einstein's equations.
\par
For a perfect fluid the assumptions of  geodesic  motion  and
nonzero rotation imply that the pressure is  constant  (see  ref.
9). Hence, from the point of view of thermodynamics, geodesic and
rotating perfect fluids are isentropic and fall within the  class
considered by Pleba\'nski [3]. However, the approach based  on  the
Darboux theorem applies to any timelike  congruence  that  is  of
class $C^{1}$ and has zero acceleration and nonzero rotation.
In particular, the velocity field of a rotating charged dust with zero Lorentz force, that was considered in several papers, has this property.
The papers that discussed such solutions will be mentioned at the end of sec. 8; they are all within the same class of the classification introduced here.
\par
In sec. 2 the Darboux theorem is introduced. In sec.  3,  the
classification of first-order differential  forms  based  on  the
Darboux  theorem  is  applied  to  geodesic  vector  fields  with
rotation. When the vector field is the velocity field of a fluid,
a class of preferred coordinates results, which shall  be  termed
"Pleba\'nski coordinates". In sec. 4, by way  of  an  example,  the
consideration of sec. 3 is applied to the rotating dust  solution
of Stephani [10]. In sec. 5 it is shown that each Killing  vector
field that might possibly exist in a rotating dust  spacetime  is
determined by two functions of  two  variables.  If  the  Killing
field is not spanned on velocity and rotation, then the Pleba\'nski
coordinates may be adapted to it so that it acquires  the  unique
form $k^{\alpha } = \delta ^{\alpha }_{1}$. The G\"odel  solution  is  used  to  illustrate  the
various forms of the Killing fields that may arise.
\par
In section 6, the consideration of secs. 3 and 5  is  applied
to the situation when there exist three  Killing  vector  fields.
When all three of them are spanned on $u^{\alpha }$ and $w^{\alpha }$,  the  result  is
trivial: the group becomes  two-dimensional,  and  this  case  is
not considered here. When two of them are spanned on $u^{\alpha }$  and $w^{\alpha }$
while the third one is not, two cases arise  that  correspond  to
different Bianchi types (II and  I)  of  the  groups.  These  are
investigated in sections 7 and 8. The solutions  of  Lanczos  [1]
and of G\"odel [2] emerge as  special  cases  in  both  types.  The
Bianchi type II metrics  are  defined  by  a  single  third-order
differential equation, the Bianchi type I metrics are  determined
by a set of autonomous first-order differential equations.
\par
Finally, in sec. 9, other invariant  definitions  are  given:
for the class III solution by Ozsvath [11] and for  the  solution
of G\"odel [2]. The former results from the assumptions:
\par
1. The source  in  the  Einstein  equations  is  a  geodesic,
rotating perfect fluid.
\par
2. The rotation vector field is covariantly constant.
\par
The G\"odel solution, which is a subcase of this, emerges  when  it
is assumed in addition that the shear of the fluid flow is zero.
\par
So far, Bianchi-type solutions of Einstein's  equations  with
rotating source have been searched for and  found  by  trial  and
error (often with nonperfect fluid sources, e.g.  with  viscosity
or heat-flow). The results of the present series of  papers  show
that, in the case of dust source at least, the number of  allowed
possibilities is limited. It  is  hoped  that  the  results  will
direct further research toward better-defined targets.
\par
\medskip
{\bf 2. The classification of differential forms  of  first  order
and the Darboux theorem.}
\par
The Darboux theorem presented below exploits the simple  fact
that if a differential form $q$ of first order is defined on an $n$-
dimensional manifold $M_{n}$, then its domain is  not  necessarily $n$-
dimensional. Two cases are well-known:
\par
1. If $q$ is a perfect differential so that $q = df$, where $f$  is
a scalar function on $M_{n}$, then $f$ can  be  chosen  as  one  of  the
coordinates, and the form becomes one-dimensional.
\par
2. If $q$ has an integrating factor so that $q$ =  {\it gdf},  where $f$
and $g$ are independent scalar functions on $M_{n}$, then $f$ and $g$ can be
chosen as two of the coordinates, and then the domain of $q$ is the
two-dimensional $(f, g)$-surface.
\par
The Darboux theorem summarizes all the cases that can  occur.
It is based on the following classification (see also Ref. 3):
\par
{\bf Definition}
\par
Let $q$ be a differential form of first order.
\par
If $Q_{2l} := dq \wedge \dots \wedge dq$ [multiplied $l$ times] $\neq  0,$ but $q \wedge  Q_{2l} = 0,$  then $q$  is
said to be of class $2l$.
\par
If $Q_{2l+1} := q \wedge  Q_{2l} \neq  0,$ but $dQ_{2l+1} \equiv  dq \wedge  Q_{2l} = 0,$ then $q$ is
said to be of class $(2l + 1)$. \square 
\par
Then the following holds:
\par
{\bf The theorem of Darboux}
\par
The form $q$ is of class $2l$ if and only if there exists  a  set
of $2l$ independent functions $(\xi _{1}, \ldots 
, \xi _{l}, \eta _{1}, \ldots 
, \eta _{l})$ such that:
\par
$$
q = \eta _{1}d\xi _{1} + \eta _{2}d\xi _{2} + \ldots 
. + \eta _{l}d\xi _{l} .
\eqnum{2.1}$$
The form $q$ is of class $(2l + 1)$ if and only if there exists a set
of $(2l + 1)$ independent functions $(\tau, \xi _{1}, \ldots 
, \xi _{l}, \eta _{1}, \ldots 
, \eta _{l})$
such that:
\par
$$
q = d\tau + \eta _{1}d\xi _{1} + \eta _{2}d\xi _{2} + \ldots 
. + \eta _{l}d\xi _{l} .
\eqnum{2.2}$$
A proof of this theorem can be found in Ref. 12.
\par
Evidently,  the  class  of $q$  cannot  be  larger  than   the
dimension of the manifold on  which $q$  is  defined.  Hence,  the
Darboux theorem implies that in a four-dimensional  spacetime $V_{4}$
any differential form of first order can be represented as:
\par
$$
q = \sigma  d\tau + \eta  d\xi ,
\eqnum{2.3}$$
where $\sigma , \tau, \eta $ and $\xi $ are scalar functions on $V_{4}$.
\par
Any vector field $u^{\alpha }$ on $V_{4}$ defines the following  differential
form:
\par
$$
q_{u} := u_{\alpha }dx^{\alpha }.
\eqnum{2.4}$$
According to (2.3), in the most general case there  exist  scalar
functions $\sigma , \tau, \eta $ and $\xi $ such that:
\par
$$
u_{\alpha } = \sigma \tau,_{\alpha } + \eta \xi ,_{\alpha }.
\eqnum{2.5}$$
Note that the functions in (2.5) are not uniquely defined.  Since
we shall not use  (2.5)  in  the  most  general  case,  we  shall
determine the nonuniqueness only in the subcase that is of direct
interest to us (see sec. 3).
\par
For the most general case of (2.5), the  four  functions  are
independent in the sense that the Jacobian:
\par
$$
{{\partial (\sigma ,\tau,\eta ,\xi )} \over {\partial (x^{0},x^{1},x^{2},x^{3})}} \neq 0.\eqnum {2.6}
$$
Hence, they can be chosen as coordinates  in  the  spacetime.  In
Refs. 4 and 7 it was shown that if $u^{\alpha }$ is the velocity  field  of
an isentropic perfect fluid  in  which  the  particle  number  is
conserved, then $\sigma  = 1/H$, where $H$ is the enthalpy per one particle
of the fluid, and further  limitations  on $u_{\alpha }$  follow  from  the
particle number conservation. No other applications of  (2.5)  in
the general case are known to this author.
\par
\medskip
{\bf 3. Geodesically moving fluids.}
\par
To any timelike vector field $u_{\alpha }$ normalized to unity (so  that
$u_{\alpha }u^{\alpha } = 1)$ the formula from Refs. 13 and 14 may be applied:
\par
$$
u_{\alpha ;\beta } = \dot{u}_{\alpha }u_{\beta } + \sigma _{\alpha \beta } + \omega _{\alpha \beta } + {1\over 3} \theta h_{\alpha \beta ,}
\eqnum{3.1}$$
which gives rise to the well-known  definitions  of  acceleration
$\dot{u}^{\alpha }$, expansion $\theta $, shear $\sigma _{\alpha \beta }$ and rotation $\omega _{\alpha \beta }$. In the signature  (+
- - -) used here, the projection tensor $h_{\alpha \beta }$ is:
\par
$$
h_{\alpha \beta } = g_{\alpha \beta } - u_{\alpha }u_{\beta }.
\eqnum{3.2}$$
The following properties of $\dot{u}^{\alpha }$, $\sigma _{\alpha \beta }$ and $\omega _{\alpha \beta }$ will be useful in
further considerations:
\par
$$
\dot{u}_{\alpha}u^{\alpha } = 0,\qquad \sigma _{\alpha \beta }u^{\beta } = \omega _{\alpha \beta }u^{\beta } = 0.
\eqnum{3.3}$$
We shall assume from now on that $u_{\alpha }$ is the velocity field  of
a fluid and that $\dot{u}_{\alpha } = 0,$ i. e. that the particles  of  the  fluid
move on geodesics. Then, from (2.5) we have:
\par
$$
\omega _{\alpha \beta } = {1\over 2} (u_{\alpha ,\beta } - u_{\beta ,\alpha }) = {1\over 2} (\sigma ,_{\beta }\tau,_{\alpha } - \sigma ,_{\alpha }\tau,_{\beta } + \eta ,_{\beta }\xi ,_{\alpha } - \eta ,_{\alpha }\xi ,_{\beta }),
\eqnum{3.4}$$
and from (3.3) we have:
\par
$$
(u^{\beta }\sigma ,_{\beta })\tau,_{\alpha } - (u^{\beta }\tau,_{\beta })\sigma ,_{\alpha } + (u^{\beta }\eta ,_{\beta })\xi ,_{\alpha } - (u^{\beta }\xi ,_{\beta })\eta ,_{\alpha } = 0.
\eqnum{3.5}$$
There are two possibilities now:
\par
I. At least one of the  four  scalar  products  in  (3.5)  is
nonzero. In this case (3.5) implies that at  most  three  of  the
functions $(\sigma , \tau, \eta , \xi )$ are independent, and  so  the  form  (2.4)
will not be of class 4.
\par
II. All the four scalar  products  are  zero.  However,  this
means that the gradients of $(\sigma , \tau, \xi , \eta )$ are all confined to  the
3-space orthogonal  to $u_{\alpha }$,  i.e.  that  there  is  a  functional
relation among these four functions. Again, the form (2.4) cannot
be of class 4.
\par
Hence, for a geodesically moving fluid the form (2.4)  is  of
class at most 3, i.e. at most 3 independent  functions $\tau, \eta , \xi $
exist such that:
\par
$$
u_{\alpha } = \tau,_{\alpha } + \eta \xi ,_{\alpha } .
\eqnum{3.6}$$
From here on, the reasoning used in Refs.  3  and  4  applies
almost unchanged. With (3.6) we have in (3.4):
\par
$$
\omega _{\alpha \beta } = {1\over 2} (\eta ,_{\beta }\xi ,_{\alpha } - \eta ,_{\alpha }\xi _{,\beta }),
\eqnum{3.7}$$
and in (3.5):
\par
$$
(u^{\beta }\eta ,_{\beta })\xi ,_{\alpha } - (u^{\beta }\xi ,_{\beta })\eta ,_{\alpha } = 0.
\eqnum{3.8}$$
There are again two possibilities:
\par
I. Either $(u^{\beta }\eta ,_{\beta })$ and $(u^{\beta }\xi ,_{\beta })$ do not  vanish  simultaneously,
and then (3.8) implies that $\eta $ and $\xi $ are functionally related,  in
which case (3.6) implies that $u_{\alpha }$ is a gradient of a function, and
so $\omega _{\alpha \beta } \equiv  0.$
\par
II. Or $\xi $ and $\eta $ are not functionally related, in which case:
\par
$$
u^{\beta }\xi ,_{\beta } = u^{\beta }\eta ,_{\beta } = 0,
\eqnum{3.9}$$
and $\omega _{\alpha \beta } \neq  0.$ We shall be interested only in the second case.  The
functions $\{\tau, \xi , \eta \}$ in (3.6) are determined up to  the  following
transformations:
\par
$$
\xi  = F(\xi ',\eta '),\qquad \eta  = G(\xi ',\eta '), \qquad \tau = \tau' - S(\xi ',\eta '), \eqnum{3.10}$$
where the functions $F$ and $G$ must obey the equation:
\par
$$
F,_{\xi '} G,_{\eta '} - F,_{\eta '} G,_{\xi '} = 1,
\eqnum{3.11}$$
and then $S$ is determined by:
\par
$$
S,_{\xi '} = GF,_{\xi '} - \eta ',\qquad S,_{\eta '} = GF,_{\eta '}.
\eqnum{3.12}$$
Eq. (3.11) is the integrability condition of eqs. (3.12)  and  it
ensures that the Jacobian of the transformation (3.10) equals  1.
It follows that one  of  the  functions $\{F, G\}$  can  be  chosen
arbitrarily, the other one is then determined by (3.11) and $S$  is
fixed  up  to  an  additive  constant  by  (3.12).  The   inverse
transformation to (3.10) is of exactly the same  form,  with  the
corresponding $F, G$ and $S$ obeying (3.11) and (3.12).
\par
Let us now make the additional assumption that the number  of
particles of the fluid is conserved, i.e.:
\par
\medskip
$$
({\sqrt{- g}} nu^{\alpha })_{,\alpha } = 0,\eqnum {3.13}$$

\noindent where $g$ is the determinant of the metric tensor and $n$ is the particle number density. This equation is a necessary and sufficient condition for the existence of a function $\zeta$ such that:
\par
\medskip
$$
{\sqrt{- g}} nu^{\alpha } = \varepsilon ^{\alpha \beta \gamma \delta }\xi ,_{\beta }\eta ,_{\gamma }\zeta ,_{\delta } .\eqnum {3.14}$$
\par
\medskip
Note that (3.6) and (3.9) imply that:
\par
$$
u^{\alpha }\tau,_{\alpha } = 1,
\eqnum{3.15}$$
and then eq. (3.14) implies that:
\par
$$
\varepsilon ^{\alpha \beta \gamma \delta }\tau,_{\alpha }\xi ,_{\beta }\eta ,_{\gamma }\zeta ,_{\delta } \equiv  {\partial (\tau,\eta ,\xi ,\zeta )\over \partial (x^{0},x^{1},x^{2},x^{3})} = \sqrt{- g} n \neq  0.
\eqnum{3.16}$$
Eq. (3.14) implies also that:
\par
$$
u^{\alpha }\zeta ,_{\alpha } = 0.
\eqnum{3.17}$$
The function $\zeta $ is determined by (3.14) up to the transformations:
\par
$$
\zeta  = \zeta ' + T(\xi ',\eta '),
\eqnum{3.18}$$
where $T$ is an arbitrary function. Eq. (3.16) certifies  that $\{\tau,
\xi , \eta , \zeta \}$ can be used as coordinates in the spacetime. If they are
chosen as the $\{x^{0}, x^{1}, x^{2}, x^{3}\} = \{t, x, y, z\}$  coordinates,
respectively, then eq. (3.6) implies:
\par
$$
u_{0} = 1, \qquad u_{1} = y, \qquad u_{2} = u_{3} = 0.
\eqnum{3.19}$$
We will use these coordinates throughout the  remaining  part  of
the paper and  call  them  "Pleba\'nski  coordinates".  Eq.  (3.16)
implies now:
\par
$$
g = -n^{-2},
\eqnum{3.20}$$
and eq. (3.14) implies:
\par
$$
u^{\alpha } = \delta ^{\alpha }_{0} ,
\eqnum{3.21}$$
i.e. the Pleba\'nski coordinates are comoving. The rotation  vector
defined by:
\par
$$
w^{\alpha } = - (1/{\sqrt{- g}}) \varepsilon ^{\alpha \beta \gamma \delta }u_{\beta }u_{\gamma ,\delta }
\eqnum{3.22}$$
assumes the form:
\par
$$
w^{\alpha } = n\delta ^{\alpha }_{3} .
\eqnum{3.23}$$
Eqs. (3.19) and (3.21) imply that:
\par
$$
g_{00} = 1,\qquad g_{01} = y,\qquad g_{02} = g_{03} = 0,
\eqnum{3.24}$$
and also that the only nonvanishing components  of  the  rotation
tensor are:
\par
$$
\omega _{12} = - \omega _{21} = 1/2.
\eqnum{3.25}$$
Note that, in contrast to Refs.  3  and  4  where  barotropic
perfect fluids were considered,  we  have  not  assumed  anything
about the form of the energy-momentum tensor so far.
\par
If we now assume that the fluid is perfect, then we  conclude
from the equations of motion ${T^{\alpha \beta }}_{;\beta } = 0$ that either $\omega  = 0$  or $p$  =
const (see also ref. 9). This means that a geodesic perfect fluid
can be rotating only if it is in fact dust; the constant $p$ can be
reinterpreted as the cosmological constant.  In  this  case,  the
energy-density obeys the conservation equation $(\sqrt{- g}  \epsilon u^{\alpha })_{,\alpha } = 0$
and eq. (3.13) need not be assumed separately\footnote 
{For dust, results closely analogous to (3.19) - (3.25) were obtained by Ellis [15], by adapting an orthonormal vector basis and  a coordinate system to $u^{\alpha}$ and $w^{\alpha}$.
Of the exact solutions with nonzero rotation found by Ellis most, but not all, do belong to the collection considered in this series of papers. They will be described in paper 2}.
\par
\medskip
{\bf 4. Example: the Stephani solution [10].}
\par
The Stephani metric with $p =$ const $\neq  0$ (eq.  (4.22)  in  Ref.
10) is not in fact a perfect fluid solution,  as  was  found  out
while trying to construct  the  Pleba\'nski  coordinates  for  it\fnote{0}{The error is deeply hidden and so far could not be corrected. I
am grateful to H. Stephani for cooperation on this point.}.
Therefore, we shall consider only the dust solution, eq. (4.8) in
Ref. 10. In the original notation except for the  signature,  the
solution is:
\par
$$
ds^{2} = \eta _{ab}dx^{a}dx^{b} - N^{2}(dx^{1})^{2} ,
\eqnum{4.1}$$
where $a, b = 0, 2, 3, \eta _{ab}$ = diag (1, -1, -1) and:
\par
$$
N := {1\over 2} M \ln  T + g_{a}x^{a} + h,
$$
$$
T^{2} := \eta _{ab}(x^{a} - f^{a}) (x^{b} - f^{b}),\eqnum {4.2}
$$
the  functions $M(x^{1}), f^{a}(x^{1}), g_{a}(x^{1})$  and $h (x^{1})$  all  being
arbitrary. The velocity field and the energy-density of the  dust
are, respectively:
\par
$$
u_{a} = T,_{a} , \qquad u_{1} = 0,
\eqnum{4.3}$$
$$
\kappa \epsilon  = M/(NT^{2}).
\eqnum{4.4}$$
The formula for the velocity field can be writen as:
\par
$$
u_{\alpha } = T,_{\alpha } - T,_{1} x^{1},_{\alpha } ,
\eqnum{4.5}$$
which  immediately  suggests  the   choice   of   the   Pleba\'nski
coordinates of sec. 3:
\par
$$
\tau = T,\qquad \xi  = x^{1},\qquad \eta  = - T,_{1}.
\eqnum{4.6}$$
Eq. (3.14) defining $\zeta $ is here:
\par
$$
MT^{-2}\eta ^{ab}T,_{b} = - \epsilon ^{a1cd}T,_{1c}\zeta ,_{d} .
\eqnum{4.7}$$
The following identity is useful in calculations:
\par
$$
{T,_{0}}^{2} - {T,_{2}}^{2} - {T,_{3}}^{2} = 1.
\eqnum{4.8}$$
Using this, one can verify that only two of the  three  equations
(4.7) are independent. The solution of (4.7) is:
\par
$$
\zeta  = M [U + \lambda (T,_{1})],
\eqnum{4.9}$$
where $\lambda (T,_{1})$ is an arbitrary function, and $U$ is defined by:
\par
$$
U = \int  [f^{3},_{1}(1 - X^{2}) - f^{0},_{1}Y(X) + f^{2},_{1}XY(X)]^{-1}dX ,
\eqnum{4.10}$$
the function $Y(X)$ being determined by:
\par
$$
T,_{1}(1 - X^{2} - Y^{2})^{1/2} = - f^{0},_{1} + f^{2},_{1}X + f^{3},_{1}\hbox{{\it Y}.\qquad }
\eqnum{4.11}$$
In the integral (4.10), the coordinate $x^{1}$ and  the  quantity $T,_{1}$
are to be treated as  parameters  independent  of  {\it X}.  After  the
integral is  calculated,  one  should  substitute  for $T,_{1}$  from
(4.11), while $X$ and $Y$ are to be replaced by:
\par
$$
X = (x^{2} - f^{2})/(x^{0} - f^{0}),\qquad Y = (x^{3} - f^{3})/(x^{0} - f^{0}),
\eqnum{4.12}$$
The integral in (4.10) is expressible in terms of  elementary
functions, but the result is rather complicated.
\par
As can be seen, the  solution  (4.1)  -  (4.4)  becomes  very
complicated in the Pleba\'nski coordinates and it is unlikely  that
it could be found in such a form  from  the  Einstein  equations.
Thus, although the coordinates are invariantly defined, they  are
not necessarily convenient for explicit calculations.
\par
A collection of other solutions represented in the  Pleba\'nski
coordinates can be found in the extended version of Ref. 6.
\par
\medskip
{\bf 5. The Killing vector fields compatible with rotation.}
\par
We shall assume that the symmetries of the spacetime (if  any
exist) are  inherited  by  the  source,  i.e.  that  if  the  Lie
derivative of the metric tensor $g_{\alpha \beta }$ along the vector field $k^{\alpha }$  is
zero, $\pounds_{k}g_{\alpha \beta } = 0,$ then the velocity field and the particle  number
density are also invariant: $\pounds_{k}u^{\alpha } = 0$ = $\pounds_{k}${\it n}. (For a  pure  perfect
fluid  source  the  inheritance  is   guaranteed.)   It   follows
that the rotation tensor must also be invariant, $\pounds_{k}\omega _{\alpha \beta } = 0.$
\par
In consequence of (3.21) the equation $\pounds_{k}u^{\alpha } \equiv  k^{\mu }u^{\alpha },_{\mu } - u^{\mu }k^{\alpha },_{\mu }
= 0$ implies that:
\par
$$
k^{\alpha },_{t} = 0.
\eqnum{5.1}$$
In consequence of (3.23) and of the assumption $\pounds_{k}n \equiv  k^{\alpha }n,_{\alpha } = 0,$
the equation $\pounds_{k}w^{\alpha } = 0$ implies:
\par
$$
k^{\alpha },_{z} = 0.
\eqnum{5.2}$$
The equation $\pounds_{k}\omega _{\alpha \beta } = 0,$ in consequence of (3.25) implies:
\par
$$
k^{1},_{x} + k^{2},_{y} = 0,
\eqnum{5.3}$$
and the equation $\pounds_{k}u_{\alpha } = 0,$ in consequence of (3.19) implies:
\par
$$
k^{0},_{x} = - k^{2} - yk^{1},_{x} ,\qquad k^{0},_{y} = - yk^{1},_{y} .
\eqnum{5.4}$$
(The equations $\pounds_{k}u_{\alpha } = 0$ and $\pounds_{k}u^{\alpha } = 0$ provide  independent  pieces
of information because the equations $\pounds_{k}g_{\alpha \beta } = 0$ have not been used
so far). Eq. (5.3) is the integrability condition of eqs.  (5.4).
The general solution of eqs. (5.1) - (5.4) is:
\par
$$
k^{0} = C + \phi  - y\phi ,_{y} , \qquad k^{1} = \phi ,_{y} , \qquad k^{2} = - \phi ,_{x} , \qquad k^{3} = \lambda ,
\eqnum{5.5}$$
where $\phi (x,y)$ and $\lambda (x,y)$ are  arbitrary  functions  and $C$  is  an
arbitrary constant. Symmetries need not be present, in  fact  the
Stephani [10] solution considered in sec. 4 is an  example  of  a
rotating dust solution with no symmetries. In this case $\phi  = \lambda  = C
= 0.$ However, if any symmetries are  present,  then  the  Killing
vector fields must have the form (5.5).
\par
Suppose that $\phi $ is not a constant, i.e. that a Killing  vector
field $k^{\alpha }$ exists that  has  a  nonzero  component  in  the $x$-  or
$y$-direction (in invariant terms this means that the vector  field
$k^{\alpha }$ is not spanned on the  vector  fields  of  velocity, $u^{\alpha }$,  and
rotation, $w^{\alpha })$. We can then, within the Pleba\'nski class defined in
sec. 3, adapt the coordinates to $k^{\alpha }$ in such  a  way  that $k^{\alpha '} =
\delta ^{\alpha '}_{1}$, i.e. so that the metric becomes  independent  of $x^\prime $.  From
(3.10) - (3.12) and (3.18) the transformation functions are:
\par
$$
t' = t - S(x,y), \qquad x' = F(x,y), \qquad y' = G(x,y), \qquad z' = z + T(x,y), \eqnum{5.6}$$
where $T$ is arbitrary, while $F, G$ and $S$ obey:
\par
$$
F,_{x}G,_{y} - F,_{y}G,_{x} = 1, \qquad S,_{x} = GF,_{x} - y,\qquad S,_{y} = GF,_{y} . \eqnum{5.7}$$
In order to lead to $k^{\alpha '} = \delta ^{\alpha '}_{1}$ the functions $F, G$ and $T$ must obey
in addition:
\par
$$
- (\phi  + C) + GF,_{x}\phi ,_{y} - GF,_{y}\phi ,_{x} = 0,
$$
$$
F,_{x}\phi ,_{y} - F,_{y}\phi ,_{x} = 1,\qquad G,_{x}\phi ,_{y} - G,_{y}\phi ,_{x} = 0,
\eqnum{5.8}$$
$$
T,_{x}\phi ,_{y} - T,_{y}\phi ,_{x} = -\lambda .
\eqnum{5.9}$$
The unique solution of eqs. (5.8) is $G = \phi  + C$, which obeys (5.7)
as well (in virtue of the second  of  (5.8)).  Eq.  (5.9)  simply
defines the accompanying $T$ which is seen to exist always. Since $\phi $
was assumed nonconstant, the transformation  is  nonsingular  (in
fact its Jacobian equals just 1), and results in $\phi  = y$ in the new
coordinates. As already noticed, the metric  becomes  independent
of $x$ after the transformation. This property is preserved by  the
transformations (5.6), but with $F, G, S$ and $T$ restricted now by:
\par
$$
G = y,\qquad F = x + H(y),\qquad T = T(y), \qquad S = \int yH,_{y}dy + A, \eqnum{5.10}$$
where  {\it A}  is  an  arbitrary  constant  and $H, T$  are  arbitrary
functions.  The  functions   given   by   (5.10)   fulfil   (5.7)
identically. Note that the transformation to $k^{\alpha '} = \delta ^{\alpha '}_{1}$  exists
irrespectively of any possible functional relation among $\lambda , \phi $ and
$C$ in (5.5); the only case when it fails is $\phi $ = const.
\par
A solution of the Einstein equations may have more  than  one
Killing vector field  of  the  form  (5.5).  In  that  case,  the
transformation (5.8) - (5.9) changes only  one  of  them  to  the
preferred form, the others will preserve their  more  complicated
appearance. An example of this situation is  the  G\"odel  solution
[2] transformed to the Pleba\'nski coordinates, see Refs. 4 and 5:
\par
$$
ds^{2} = (dt + y dx)^{2} - {1\over 2} y^{2}dx^{2} - (\kappa \epsilon y^{2})^{-1}dy^{2} - 2\kappa \epsilon ^{-1}dz^{2},
\eqnum{5.11}$$
where $\kappa  = 8\pi G/c^{4}$ and $\epsilon $  is  the  energy-density  related  to  the
cosmological constant by\fnote{1}{Note that if the cosmological constant is reinterpreted as pressure, then the resulting perfect fluid has the equation of state
$\epsilon  = p$. Hence, the G\"odel solution may have  been  the  first
example considered in the literature of a "stiff perfect  fluid",
now familiar from the studies of solution-generating  techniques, see e.g. Verdaguer [16].}:
$$
\Lambda  = {1\over 2} \kappa \epsilon .
\eqnum{5.12}$$
The  symmetry  group  of  this  solution  is  5-dimensional,  the
independent 1-parametric subgroups were given in  Ref.  4.  Those
connected with nonconstant $\phi $ in (5.5) are the following three:
\par
$$
x = x' + s_{1},
\eqnum{5.13}$$
$$
x = e^{-s_{2}}x',\qquad y = e^{s_{2}}y',
\eqnum{5.14}$$
\par
$$t = t' + (2\sqrt{2}/K) \arctan  [\sqrt{2} s_{3}(Ky')^{-1}(1 - s_{3}x')^{-1}],$$ 
$$
x = [K^{2}\hbox{{\it x}'{\it y}'}^{2}(1 - s_{3}x') - 2s_{3}] / [2s^{2}_{3} + K^{2}y'^{2}(1 - s_{3}x')^{2}],
$$
$$
y = (1 - s_{3}x')^{2}y' + 2s^{2}_{3}/(K^{2}y'),
\eqnum{5.15}$$
where $s_{1}, s_{2}$ and $s_{3}$ are the group parameters and $K := (\kappa \epsilon )^{1/2}$.
The Killing vectors are, respectively: $k_{(1)}^{\alpha } = \delta ^{\alpha }_{1}$ (corresponding
to $C = \lambda  = 0, \phi  = y$, the one constructed in (5.8) - (5.9)), $k_{(2)}^{\alpha } =
- x\delta ^{\alpha }_{1} + y\delta ^{\alpha }_{2}$  (corresponding to $\phi  = - xy$) and $k_{(3)}^{\alpha } = 4(K^{2}y)^{-1}\delta ^{\alpha }_{0} +
[x^{2} - 2/(Ky)^{2}]\delta ^{\alpha }_{1} - 2xy\delta ^{\alpha }_{2}$ (corresponding to $\phi  = x^{2}y + 2/(K^{2}y)$).
\par
\medskip
{\bf 6. The algebra of three Killing fields.}
\par
Suppose that three Killing vector fields exist and all  three
are spanned on $u^{\alpha }$ and $w^{\alpha }$, so that $\phi $ = const in (5.5) for each  of
them, i.e.:
\par
$$
k_{(i)}^{\alpha } = C_{i}\delta ^{\alpha }_{0} + \lambda _{i}(x, y) \delta ^{\alpha }_{3}, \qquad i = 1, 2, 3.
\eqnum{6.1}$$
From the Killing equations one can then easily conclude that constants $\alpha_1$, $\alpha_2$ and $\alpha_3$ exist such that $\alpha_1k_{(1)} + \alpha_2k_{(2)} + \alpha_3k_{(3)} = 0$, i.e. the symmetry group is  in  fact  two-dimensional.  Hence,  no  three-
dimensional symmetry group with the generators (6.1) exists\fnote{2}{The algebra (6.1) corresponds to a three-dimensional group  that
has two-dimensional orbits, and it turns out  that  in  the  case
considered the group has to be two-dimensional as well.  As  will
follow from the whole of the present  work,  three-dimensional  symmetry
groups with two-dimensional orbits just do not exist for rotating
dust.}; for
a three-dimensional group at least one of the generators must  be
linearly independent of $u^{\alpha }$ and $w^{\alpha }$ at every point of the spacetime region under consideration.
\par
In sections 7 and 8 we  shall  consider  the  situation  when
exactly one generator, $k_{(1)}^{\alpha }$, is everywhere linearly independent of $u^{\alpha }$ and $w^{\alpha }$,
while the other two, $k_{(2)}^{\alpha }$ and $k_{(3)}^{\alpha }$,  are  of  the  form  (6.1).  In
agreement with the result of sec. 5,  the  Pleba\'nski  coordinates
can be adapted to $k_{(1)}^{\alpha }$ so that:
\par
$$
k_{(1)}^{\alpha } = \delta^{\alpha }_{1},\eqnum {6.2}$$
\medskip
while:
\par
$$
k_{(2)}^{\alpha } = C_{2}\delta ^{\alpha }_{0} + \lambda _{2}(x, y)\delta ^{\alpha }_{3} , \qquad k_{(3)}^{\alpha } = C_{3}\delta ^{\alpha }_{0} + \lambda _{3}(x, y)\delta ^{\alpha }_{3} ,\eqnum{6.3}
$$
and the coordinate transformations preserving (6.2) and (6.3) are
(5.10). Note that $C_{2}$ and $C_{3}$ cannot vanish simultaneously  because
otherwise the Killing equations immediately imply that either $k_{(3)}^{\alpha } =$ const $k_{(2)}^{\alpha }$ (in which case the symmetry group is  two-dimensional)
or the metric is singular. However, with no loss of generality we
can assume that:
\par
$$
C_{2} \neq  0 = C_{3}
\eqnum{6.4}$$
because the Killing vector fields are  determined  up  to  linear
combinations among them. Hence, if initially $C_{2} \neq  0 \neq  C_{3}$, then we
take $k_{(3)}'^{\alpha } = k_{(3)}^{\alpha } - (C_{3}/C_{2}) k_{(2)}^{\alpha }$ instead of $k_{(3)}^{\alpha }$ as the basis  generator.
If initially $C_{2} = 0 \neq  C_{3}$, then we exchange  the  labels  "2"  and
"3".
\par
We shall keep the choice (6.4) throughout sections 7 and 8.
\par
\medskip
{\bf 7. The generators, the Killing  equations  and  the  Einstein
equations for a Bianchi type II algebra.}
\par
All the cases that arise follow as limits at different stages
of calculation from the  generic  case $\lambda _{2} \neq  0,$  and  we  shall
consider it first. The commutators of the Killing vectors are:
\par
\medskip
$$
[k_{(1)} , k_{(2)}]^{\alpha } = (\lambda _{2,x} /\lambda _{3}) k_{(3)}^{\alpha },\qquad [k_{(2)} , k_{(3)}]^{\alpha } = 0,$$
\par
\medskip
$$
[k_{(1)} , k_{(3)}]^{\alpha } = (\lambda _{3,x} /\lambda _{3}) k_{(3)}^{\alpha } .\eqnum {7.1}$$
\medskip
The Killing vector fields will thus form a Lie algebra when:
\par
$$
\lambda _{2,x} = b\lambda _{3} , \qquad \lambda _{3,x} = c\lambda _{3} ,
\eqnum{7.2}$$
where $b$ and $c$ are arbitrary constants. The case $c \neq 0$ has to  be
considered separately. Then:
\par
$$
\lambda _{3} = \beta (y) e^{cx},\qquad \lambda _{2} = (b/c) \beta (y) e^{cx} + \alpha (y),
\eqnum{7.3}$$
where $\alpha (y)$ and $\beta (y)$ are arbitrary  functions.  However,  in  this
case we can take $k_{(2)}'^{\alpha } = k_{(2)}^{\alpha } - (b/c) k_{(3)}^{\alpha }$ as the new  basis  generator
instead of $k_{(2)}^{\alpha }$, and the result is equivalent to assuming $b = 0.$
Hence, with $c \neq 0$, we can take $b = 0$ with no loss of generality.
\par
The further procedure consists of the following steps:
\par
1. Adapt the coordinates to the Killing fields to  make  them
as simple as possible.
\par
2. Solve the Killing equations  for  the  components  of  the
metric tensor.
\par
3. Simplify the metric  as  far  as  possible  by  coordinate
transformations.
\par
4. Solve the Einstein equations.
\par
The coordinate transformations in steps 1 and 3 in  general  lead
out of the Pleba\'nski class defined in sec. 3.
\par
This procedure will be presented in some detail below. In the
present case the result from the Einstein equations is: either $c
= 0$ or there is no rotation. Since we are interested in  rotating
solutions only, the case $c \neq  0 = b$ need not be followed  further.
We thus assume:
\par
$$
c = 0.
\eqnum{7.4}$$
Then:
\par
$$
\lambda _{3} = \beta (y),\qquad \lambda _{2} = b\beta (y) x + \alpha (y).
\eqnum{7.5}$$
The algebra of the Killing vector fields is of  Bianchi  type  II
when $b \neq  0$ and of Bianchi type I when $b = 0.$
\par
In order to simplify the Killing vectors we now transform the
coordinates as follows:
\par
$$
(t', x', y') = (t, x, y), \qquad z' = - (\alpha /C_{2})t + z/\beta .
\eqnum{7.6}$$
The transformation  is  not  of  the  form  (5.10),  so  the  new
coordinates do not belong to the Pleba\'nski class, and  the  forms
of velocity, rotation and the metric will no  longer  agree  with
(3.19) - (3.25). The Killing vector fields in the new coordinates
become:
\par
$$
k_{(1)}^{\alpha } = \delta ^{\alpha }_{1},\qquad k_{(2)}^{\alpha } = \delta ^{\alpha }_{0} + bx\delta ^{\alpha }_{3}, \qquad k_{(3)}^{\alpha } = \delta ^{\alpha }_{3} ,
\eqnum{7.7}$$
while the velocity and rotation fields become:
\par
$$
u^{\alpha } = \delta ^{\alpha }_{0} - (\alpha /C_{2})\delta ^{\alpha }_{3}, \qquad w^{\alpha } = (n/\beta )\delta ^{\alpha }_{3}.
\eqnum{7.8}$$
The transformed metric is independent  of $x$  and $z$,  while  the
Killing equations for $k_{(2)}^{\alpha }$ imply:
\par
$$
g_{00} = 1 + (\alpha /C_{2})^{2}h_{33}(y), \qquad g_{01} = y + (\alpha /C_{2})g_{13} ,
$$
$$
g_{02} = 0, \qquad g_{03} = (\alpha /C_{2})h_{33}(y),
$$
$$
g_{11} = h_{33}(y)(bt)^{2} - 2 h_{13}(y) bt + h_{11}(y),
$$
$$
g_{12} = - h_{23}(y) bt + h_{12}(y),\qquad g_{13} = - h_{33}(y) bt + h_{13}(y),
$$
$$
g_{22} = h_{22}(y),\qquad g_{23} = h_{23}(y),\qquad g_{33} = h_{33}(y),
\eqnum{7.9}$$
where $h_{ij}(y), i, j = 1, 2, 3,$ are arbitrary functions of $y$, to be
found from the Einstein equations.
\par
The orbits of the symmetry group are now the hypersurfaces $y = {\rm const}$. In order to follow the  standard
technique of the Bianchi-type spaces we should now  carry  out  a
coordinate transformation that  preserves  (7.7)  and  makes  the
$y$-coordinate curves orthogonal to the group orbits, so that $g'_{02}
= g'_{12}= g'_{23} = 0$ after the transformation. This step  is  not  in
fact necessary for solving the Einstein equations (in general  it
only reshuffles the unknown functions without elliminating any of
them), but  in  the  case  under  consideration  it  leads  to  a
simplification. The transformation is:
\par
$$
t = t' + f_{0}(y'), \qquad x = x' + f_{1}(y'), \qquad y = f_{2}(y'),
$$
$$
z = z' + bf_{1}(y')t' + f_{3}(y'),
\eqnum{7.10}$$
where $f_{\alpha }(y')$ must obey:
\par
$$
f_{0,y'} = - f_{2}f_{1,y'} + (\alpha /C_{2})h_{23} f_{2,y'} ,
$$
$$
(- f^{2}_{2} + h_{11} - h^{2}_{13}/h_{33}) f_{1,y'} + (h_{12} - h_{13}h_{23}/h_{33} + \alpha f_{2}h_{23}/C_{2}) f_{2,y'} = 0,
$$
$$
f_{3,y'} = - (\alpha /C_{2})f_{0,y'} + (bf_{0} - h_{13}/h_{33})f_{1,y'} - (h_{23}/h_{33})f_{2,y'}. \eqnum{7.11}
$$
The equations are well-defined because:
\par
I. $h_{33} \neq  0$; otherwise the  rotation  vector  would  be  null,
which is a physical impossibility.
\par
II. $-f^{2}_{2} + h_{11} - h^{2}_{13}/h_{33} \neq  0$; otherwise the  determinant  of
the metric tensor becomes positive; i.e. the metric  acquires  an
unphysical signature.
\par
Eqs. (7.11) are to be understood  as  follows.  The  function
$f_{2}(y')$ can be chosen arbitrarily, therefore we choose it so  that
$g'_{22} = -1$ after the  transformation.  With $f_{2}(y')$  thus  chosen,
$f_{1}(y')$ is found from the second of (7.11), then $f_{0}(y')$  is  found
from the first of (7.11), and finally $f_{3}(y')$ is  found  from  the
third of (7.11).
\par
After the transformation the metric becomes (primes dropped):
\par
$$
g_{00} = 1 + (\alpha /C_{2} + bf_{1})^{2}h_{33} ,
$$
$$
g_{01} = Y(y) + (\alpha /C_{2} + bf_{1}) (- h_{33} bt + h_{13} - bf_{0}h_{33}),
$$
$$
g_{02} = g_{12}= g_{23} = 0,
$$
$$g_{03} = (\alpha /C_{2} + bf_{1}) h_{33},$$ 
$$
g_{11} = h_{33} b^{2}(t + f_{0})^{2} - 2 h_{13} b(t + f_{0}) + h_{11} ,
$$
$$
g_{13} = - h_{33} b(t + f_{0}) + h_{13} ,
$$
$$
g_{22} = -1, \qquad g_{33} = h_{33}(y),
\eqnum{7.12}$$
where $h_{ij}(y), i, j = 1, 2, 3, f_{0}(y), f_{1}(y), Y(y) = f_{2}(y)$ and $\alpha (y)$
are functions to be found from the Einstein equations, and $b$  and
$C_{2}$ are arbitrary constants, $C_{2} \neq  0.$
\par
For  convenience  in  calculations  we  introduce   the   new
functions $G(y), A(y), k_{13}(y)$ and $F(y)$ by:
\par
$$
g_{33} = - G^{2}(y),\qquad \alpha /C_{2} + bf_{1} = A(y), \qquad h_{13} = - G^{2}(k_{13} + bf_{0}),
$$
$$
h_{11} = Y^{2} - F^{2} - k^{2}_{13}G^{2} + b^{2}G^{2}f^{2}_{0} - 2bf_{0}k_{13}G^{2}. \eqnum{7.13}
$$
The velocity field in the coordinates of (7.12) - (7.13) is:
\par
$$
u^{\alpha } = \delta ^{\alpha }_{0} - A\delta ^{\alpha }_{3}.
\eqnum{7.14}$$
Now the metric form is:
\par
$$
ds^{2} = (dt + Y dx)^{2} - (F dx)^{2} - dy^{2} - G^{2}[A dt - (bt - k_{13}) dx + dz]^{2}.
\eqnum{7.15}$$
The components of the Einstein tensor will  be  referred  to  the
orthonormal tetrad of forms $e^{i} = e^{i}_{\alpha } dx^{\alpha }, i = 0, 1, 2, 3,$
uniquely implied by (7.15). Note that $e^{0} = u_{\alpha } dx^{\alpha }$.  Hence,  the
Einstein equations are:
\par
$$
G_{00} = (8\pi G/c^{4}) \epsilon  ,
$$
$$
G_{11} = G_{22} = G_{33} = \Lambda ,\qquad G_{ij} = 0\hbox{ when }i \neq  j,
\eqnum{7.16}$$
where $\epsilon $ is the energy-density and $\Lambda $ is the cosmological constant.
\par
The equation $G_{12} = 0$ implies that $bA,_{y} = 0.$ The case $b = 0$
will be considered separately below, so we take here:
\par
\medskip
$$A = {\rm const}. \eqnum {7.17}$$
\medskip
Then $G_{02} = 0$ implies:
\medskip
$$k_{13} = {\rm const}. \eqnum {7.18}$$
\medskip
We can then carry out the coordinate transformation:
\par
$$
z = z' - At - k_{13}x,\qquad (t, x, y) = (t', x', y'),
\eqnum{7.19}$$
which has the same result as if:
\par
$$
A = k_{13} = 0,
\eqnum{7.20}$$
and we shall assume  this  from  now  on.  The  metric  is  still
independent of $x$ and of $z$, while $A = k_{13} = 0$  implies $g_{13} = 0,$
i.e. the Killing vectors $k_{(1)}^{\alpha } = \delta ^{\alpha }_{1}$ and $k_{(3)}^{\alpha } = \delta ^{\alpha }_{3}$ are orthogonal  to
each other. The equation $G_{01} = 0$ then has the integral:
\par
$$
Y,_{y} G/F = B =\hbox{ const},
\eqnum{7.21}$$
and we can assume $B \neq  0$ because rotation would be zero with $B = 0
= Y,_{y}$.
\par
At this point, only the diagonal components of  the  Einstein
tensor are still nonzero, of which $G_{00}$ just defines  the  energy-
density, and the other three are functionally dependent (i.e.  if
$G_{11}= \Lambda  = G_{22}$ are fulfilled, then so is $G_{33} = \Lambda )$.  They  determine
$F(y)$ and $G(y)$.
\par
It is convenient to introduce $Y(y)$ as the new  variable.  The
equation $G_{11} + G_{22} = 2\Lambda $ can then be written,  with  the  help  of
(7.21), as:
\par
$$
(F^{2}G,_{Y}/G),_{Y} = 2\Lambda G^{2}/B^{2} - {1\over 2} ,
\eqnum{7.22}$$
and so:
\par
$$
F^{2} = (C - {1\over 2} Y + 2 {\Lambda \over B^{2}} \int  G^{2}dY) G/G,_{Y} ,
\eqnum{7.23}$$
where $C$ is a new arbitrary  constant  (we  can  assume $G,_{Y} \neq  0$
because $G,_{Y} = 0$ implies $b = 0$ from $G_{11} - G_{22} = 0,$ and $b = 0$  will
be considered separately). Using (7.23) in $G_{22} = \Lambda $ we obtain  the
following integro-differential equation that determines $G$:
\par
$$
- {1\over 4} b^{2}GG,_{Y} + {1\over 2} (B/G)^{2}(C - {1\over 2} Y + 2 {\Lambda \over B^{2}} \int  G^{2}dY)^{2}(G,_{Y}/G - G,_{YY}/G,_{Y}) = 0. \eqnum{7.24}$$
\par
\medskip
In the special case $\Lambda  = 0$ this becomes an  ordinary  second-order
differential equation. It is easy to get rid of the  integral  by
transforming (7.24) appropriately and differentiating the  result
by $Y ($in this way a third-order differential equation for $G(Y)$ is
obtained) or by introducing the new variable $u(Y)$ by $dY/du = 1/G^{2}$ 
(this results in a second-order equation for $G(u)$).  However,  no
progress toward solving (7.24) results in either case.
\par
With the help of the equations $G_{11} = \Lambda  = G_{22}$ the formula  for
energy-density may be simplified to:
\par
$$
(8\pi G/c^{4})\epsilon  = (B/G)^{2} - (bG)^{2} - 2\Lambda .
\eqnum{7.25}$$
Note that the solutions considered here have a meaningful limit $b = 0.$ 

When $G$ = const, eqs. (7.23) and (7.24) no longer  apply  and
one has to go back to the Einstein equations. They imply:
\par
$$
G^{2} = B^{2}/(4\Lambda )
\eqnum{7.26}$$
(i.e. necessarily $\Lambda  > 0)$ and:
\par
$$
F^{2} = {1\over 2} Y^{2} + DY + E,
\eqnum{7.27}$$
where $D$ and $E$ are constants. If $Y$ is chosen as the new coordinate
in place of $y$, then from (7.21) and (7.26) the  metric  component
$g_{YY}$ is:
\par
$$
g_{YY} = - (G/BF)^{2} = -1/(4\Lambda F^{2})
\eqnum{7.28}$$
and the resulting metric is the G\"odel solution (see Ref. 4). Note
that $G$ = const is equivalent to $\epsilon $ = const, see eq. (7.25).
\par
When $G,_{Y} \neq  0 = b$, eq. (7.24) implies $G = e^{DY+E}$,  and  this
leads to the Lanczos solution (see Ref. 4).
\par
These derivations of the Lanczos and G\"odel solutions lead  to
their invariant definitions that are based on weaker  assumptions
than the definitions known so far:
\par
1. The source in the Einstein equations is a rotating dust.
\par
2. The spacetime has a 3-dimensional symmetry group.
\par
3. Two of the symmetry generators are spanned on  the  vector
fields of velocity $u^{\alpha }$ and rotation $w^{\alpha }$, while  the  third  one  is
linearly independent of $u^{\alpha }$ and $w^{\alpha }$ at every point.
\par
4. The generators form a Bianchi type II algebra.
\par
5. In the solutions of the Einstein  equations,  the  Bianchi
type I limit is taken of the Bianchi type II symmetry.
\par
6. The G\"odel solution  results  when  the  matter-density  is
constant, the Lanczos solution results when the  density  is  not
constant.
\par
The generalization with respect to the earlier definition  is
contained in point 3: in previous derivations the two  generators
were assumed to be collinear with $u^{\alpha }$ and $w^{\alpha }$,  respectively,  from
the beginning.
\par
\medskip
{\bf 8. The generators, the Killing  equations  and  the  Einstein
equations for a Bianchi type I algebra.}
\par
We shall consider the case $b = c = 0$ in (7.1)  -  (7.2).  The
reasoning up to eq. (7.16)  applies  also  here,  but  (7.17)  no
longer follows. Instead, the equation $G_{13} = 0$ can  be  integrated
with the result:
\par
$$
k_{13,y} = BF/G^{3} - YA,_{y} ,
\eqnum{8.1}$$
where $B$ is an arbitrary constant;  the  equation $G_{01}$  =  can  be
integrated to:
\par
$$
Y,_{y} = (C - BA)F/G,
\eqnum{8.2}$$
where $C$ is an arbitrary constant; and the equation $G_{03} = 0$ can be
integrated to:
\par
$$
A,_{y} = (\hbox{{\it BY} }- D)/(FG^{3}),
\eqnum{8.3}$$
where $D$ is one more arbitrary constant.
\par
At this point, only the diagonal components of  the  Einstein
tensor survive, and $G_{00} = (8\pi G/c^{4})\epsilon  - \Lambda $ just defines the  energy-
density. The equations $G_{11} = \Lambda  = G_{22} = G_{33}$ can be written as\fnote{4}{In order to arrive at this form, one has  to  calculate $B$  from
(8.1) and replace one factor $B$ in $B^{2}$ by the resulting expression;
then replace one $Y,_{y}$ in $Y,^{2}_{y}$ from (8.2) and replace  one $A,_{y}$  in
$A,^{2}_{y}$ from (8.3).
\par
}
:
\par
$$
- {B\over 4FG} k_{13,y} + {C - BA\over 4FG} Y,_{y} + G,_{yy} /G - {2\hbox{{\it BY} }- D\over 4FG} A,_{y} = \Lambda ,
\eqnum{8.4}$$
$$
- {B\over 4FG} k_{13,y} + {C - BA\over 4FG} Y,_{y} + {F,_{y}G,_{y}\over FG} - {D\over 4FG} A,_{y} = \Lambda ,
\eqnum{8.5}$$
$$
{3B\over 4FG} k_{13,y} - {C - BA\over 4FG} Y,_{y} + F,_{yy} /F +{3D\over 4FG} A,_{y} = \Lambda .
\eqnum{8.6}$$

\medskip

\noindent The set (8.4) - (8.6) can be integrated to a first order set. Subtracting (8.6) from (8.4) and multiplying the result by $FG$ we obtain an equation that is easily integrated to:
\par
$$
FG,_{y} - GF,_{y} - Bk_{13} - {1\over 2}\hbox{ {\it BAY} }+ {1\over 2} CY - {1\over 2} DA = E =\hbox{ const.\qquad }
\eqnum{8.7}$$
Now adding (8.3) and (8.4), and multiplying the result by $FG$  we
obtain another integrable equation whose integral can be  written
in the form:
\par
$$
FG,_{y} = {1\over 2} Bk_{13} + {1\over 2}\hbox{ {\it BAY} }- {1\over 2} CY + 2\Lambda \int FG dy + H_{0} ,
\eqnum{8.8}$$
where $H_{0}$ is an arbitrary constant. The integral can be calculated
if the new variable $u(y)$ is introduced by:
\par
$$
dy/du = 1/(FG).
\eqnum{8.9}$$
From (8.7) and (8.8) it follows that:
\par
$$
GF,_{y} = - {1\over 2} Bk_{13} - {1\over 2} DA - E + 2\Lambda \int FG dy + H_{0} .
\eqnum{8.10}$$
In the set (8.4) - (8.6) there remains one equation that  has
still not been used. However, at this point it merely  introduces
a relation  between  the  arbitrary  constants,  i.e.  implicitly
defines $H_{0}$ in terms of the  other  constants.  This  is  seen  as
follows: substitute for $k_{13,y}, Y,_{y}, F,_{y}, G,_{y}$ and $A,_{y}$ from (8.1) -
(8.3), (8.8) and (8.10) in (8.5), thereby obtaining an  algebraic
equation (i.e. one without derivatives). Differentiate  it  by $y$
and elliminate the  derivatives  in  the  same  way  again.  What
results is an identity 0 = 0. Hence, the left-hand side of  (8.5)
is identically constant in virtue of the other equations.
\par
In terms of the variable $u$ from (8.9), eqs.  (8.1)  -  (8.3),
(8.8) and (8.10) form an autonomous set of first-order  equations
that can be investigated further by qualitative methods (see e.g.
Ref. 17). This is left as a subject for a separate study.
\par
In analogy with the  Bianchi  type  I  spatially  homogeneous
(nonrotating) dust solutions (see Ref. 18) one might expect
further progress by adapting the  coordinates  suitably  (in  the
case considered in Ref. 18,  the  metric  can  be  diagonalized).
However, this author was not able to achieve any such progress.
\par
The functions {\it A}({\it y}) and $k_{13}(y)$ have  invariant  meaning:  they
are proportional to the scalar products of  the  Killing  vectors
(see eqs. (7.7) and (7.15) with $b = 0)$:
\par
$$
A = - g_{\alpha \beta }k_{(2)}^{\alpha }k_{(3)}^{\beta }/G^{2},\qquad k_{13} = - g_{\alpha \beta }k_{(1)}^{\alpha }k_{(3)}^{\beta }/G^{2}
\eqnum{8.11}$$
(note that $G^{2} = - g_{\alpha \beta } k_{(3)}^{\alpha }k_{(3)}^{\beta }$, i.e. it is a scalar, too). Hence, {\it A} =
0 and $k_{13} = 0$ are invariant properties. Note that {\it A} = 0  implies,
through (8.3), that either $Y$ = const (in which case there  is  no
rotation) or $B = D = 0.$ In the latter case, $k_{13}$ = const  and  the
coordinate transformation $z = z' - k_{13}x$ leads to $k_{13} = 0$  in  the
new coordinates. With $A = k_{13} = 0,$ the Lanczos and  G\"odel  models
result from the Einstein equations as the only solutions.  Hence,
one more invariant definition of these models follows, similar to
the six-point definition at the end of section 7. Points 1, 2,  3
and 6 remain unchanged, while points 4 and 5 are replaced by:
\par
$4'$. The generators form a Bianchi type I algebra.
\par
$5'$. From the two generators spanned on $u^{\alpha }$ and $w^{\alpha }$, two  linear
combinations can be  constructed  that  are  orthogonal  to  each
other.
\par
Point $5'$ is equivalent to the  existence  of  coordinates  in
which {\it A} = 0.
\par
Note that the  Bianchi  type  I  models  considered  in  this
section are more general than the Bianchi type  I  limit  of  the
models from sec. 7; those from sec. 7 had $A = k_{13} = 0$  in  virtue
of Einstein's equations.
\par
The assumption $k_{13} = 0$ (i.e. $g_{\alpha \beta }k_{(1)}^{\alpha }k_{(3)}^{\beta } = 0$)  alone  does  not
lead to any immediate progress in solving the Einstein equations.

The Lanczos solution was originally derived in Ref. 1 (an English translation, Ref. 19, is now available), and rediscovered in Ref. 20.
Its limit of zero cosmological constant was rediscovered in Ref. 21 as the cylindrically symmetric subcase of a family of stationary axially symmetric solutions. Geometrical and physical properties of the Lanczos solution were discussed in Ref. 1, and, in a more modern language, also in Ref. 22 (the latter only for the case $\Lambda = 0$).

Coordinate transforms of the G\"{o}del solution were published as new solutions in Refs. 23 and 24 (concerning Ref. 23 see also Ref. 25).

A metric form that is a modest generalization of the G\"{o}del solution (it has two unknown functions of one variable in place of G\"{o}del's ${\rm e}^{x^1}$ and ${\rm e}^{2x^1}$) came to be known as "G\"{o}del-type metric" and became the subject of a rather large number of papers; the activity seems to have started with Ref. 26, one of the most recent appearances of it is Ref. 27. However, it was proven already in Ref. 28 that the only perfect fluid solution with this metric is the G\"{o}del solution itself; indeed, all other "G\"{o}del-type solutions" have various nonperfect fluid sources, and therefore they do not show up in the scheme considered here.

As mentioned in sec. 1, several authors considered rotating charged dust solutions under the additional assumption that  the electromagnetic field $F_{\mu \nu}$ exerts no force on the charged dust particles, i.e. that $F_{\mu \nu}u^{\nu} = 0$. 
These solutions were all derived with another, rather natural assumption: that all charges are attached to dust particles so that no currents are present apart from the one created by the dust flow. Those solutions are found in Refs. 29 - 36.
The one in Ref. 29 has only two-dimensional symmetry, so it could not come up in this investigation.
The remaining ones are stationary and cylindrically symmetric and would have shown up here, had we allowed charges and electromagnetic fields in the source. They have the following properties:

The one from Ref. 30 becomes a vacuum solution in the limit $F_{\mu \nu} = 0$.

The one from Ref. 31 does not allow this limit at all.

The limit $F_{\mu \nu} = 0$ of the solution from Ref. 32 is the Minkowski metric.

The Som - Raychaudhuri solution [33] reproduces the $\Lambda = 0$ subcase of the Lanczos solution when $F_{\mu \nu} = 0$.

The first of the six solutions by Banerjee and Banerji [34] reduces to the G\"{o}del solution when $F_{\mu \nu} = 0$.
 The other five behave as follows: 2 and 6 become vacuum solutions when $F_{\mu \nu} = 0$, no 5 becomes the Minkowski spacetime, no 3 does not allow this limit at all, and no 4 has has a two-dimensional symmetry group.

Both solutions by Mitski\'{e}vi\v{c} and Tsalakou [35] are generalizations of the G\"{o}del solution; the first one of them is in addition a generalization of the full ($\Lambda \neq 0$) Lanczos solution\footnote{In fact, the second solution has nonzero pressure gradient that remains nonzero even after the limit $F_{\mu \nu} = 0$ is taken. Another limiting transition, given in the paper, reduces the solution to G\"{o}del's.}. In the limit $\Lambda = 0$, the first solution reduces to the one by Som and Raychaudhuri [33].

The two solutions from Ref. 36 are coordinate transforms of those from Ref. 35.

Three other generalizations of the G\"{o}del solution exist in the literature that have zero acceleration. Two were provided by Raval and Vaidya [37]; the first of them is stationary, the second expanding, both have anisotropic pressure. The third is the solution by Rebou\c{c}as [38] in which the source is a free electromagnetic field (see also Ref. 39). 
The metric of the Rebou\c{c}as solution is the same as that in the first Banerjee - Banerji solution. This coincidence was explained by Raychaudhuri and Guha Thakurta [40]: The two electromagnetic fields (one generated by a current, the other source-free) are related by a point-dependent duality rotation.
\par
\medskip
{\bf 9. Another invariant definition  of  the  G\"odel  and  Ozsvath
class III solutions.}
\par
Assumptions about invariant properties of the velocity  field
of matter usually  lead  to  progress  in  solving  the  Einstein
equations; the most impressive example were the shearfree  normal
models of Barnes [41], where a large class of solutions  resulted
from the assumptions of zero shear and zero rotation in a perfect
fluid source. Inspired by this, one can try to  make  assumptions
about other vector fields characterizing fluid sources, e.g.  the
rotation. Indeed, it turns out that the assumption:
\par
$$
w_{\alpha ;\beta } = 0,
\eqnum{9.1}$$
i.e. the rotation field being covariantly constant, together with
the assumption of geodesic motion  of  a  perfect  fluid  source,
leads uniquely to two solutions of Einstein's equations. However,
both of them were obtained before by other methods.  One  is  the
Ozsvath class III metric [11], originally identified  as  one  of
the solutions that are homogeneous in four dimensions; the  other
is the G\"odel solution [2] which is the  shearfree  limit  of  the
Ozsvath solution.
\par
From (9.1) and from the Ricci identity $2w_{\alpha ;[\beta \gamma ]} = {R^{\rho }} _{\alpha \beta \gamma } w_{\rho }$
one obtains for the Ricci tensor:
\par
$$
{R^{\rho }} _{\gamma } w_{\rho } = 0,
\eqnum{9.2}$$
and then from the Einstein equations for a perfect fluid:
\par
$$
G_{\alpha \beta } + \Lambda g_{\alpha \beta } = \kappa  [(\epsilon  + p)u_{\alpha }u_{\beta } - pg_{\alpha \beta }],\qquad \kappa  = 8\pi G/c^{4},
\eqnum{9.3}$$
and from $u^{\alpha }w_{\alpha } = 0$ one obtains:
\par
$$
\Lambda  = {1\over 2} \kappa  (\epsilon  - p).
\eqnum{9.4}$$
In the case $\Lambda  = 0,$ this is the well-known "stiff perfect  fluid".
Eq. (9.4) is a necessary condition for (9.1) when the source is a
perfect fluid.
\par
As stated at the end of sec. 3, if the  perfect  fluid  moves
geodesically with rotation, then necessarily $p$ = const. Eq. (9.4)
implies then $\epsilon $ = const, i.e. a geodesically moving  and  rotating
perfect fluid whose rotation vector is covariantly constant  must
have constant matter density. Since $\Lambda  = 0$ may be assumed with  no
loss of generality (this leads only to redefining $p)$,  we  shall
assume this from now on. Then $\epsilon  = p$ and $\epsilon  + p = 2p$ is a conserved
quantity. Hence, we may assume:
\medskip
$$n = \epsilon  + p = 2p = {\rm const} \eqnum {9.5}$$
\medskip
in all formulae. In particular, (3.20) implies then:
\par
$$
g = \det (g_{\alpha \beta }) = -\hbox{ {\it hlA}}^{2} = - (2p)^{-2}.
\eqnum{9.6}$$
Using (9.6) and (3.23) in  (9.1)  we  obtain,  in  the  Pleba\'nski
coordinates:
\par
$$
{1\over 2} n (g_{\alpha 3,\beta } - g_{\beta 3,\alpha } + g_{\alpha \beta ,3}) = 0.
\eqnum{9.7}$$
After a simple  algebraic  manipulation  this  set  of  equations
yields the following result:
\par
$$
g_{33} = - A^{2} =\hbox{ const},\qquad g_{\alpha \beta ,z} = 0,\qquad g_{13,t} = g_{23,t} = 0,
$$
$$
g_{23,x} - g_{13,y} = 0.
\eqnum{9.8}$$
The second equation in (9.8) means that $w^{\alpha }$ is a  Killing  vector,
as should be expected from (9.1), (9.6) and  (3.23).  Eqs.  (9.8)
imply that $g_{13}$ and $g_{23}$ depend only on $x$ and $y$,  and  that  there
exists a function ${\cal F}(x, y)$ such that:
\par
$$
g_{23} = {\cal F},_{y} ,\qquad g_{13} = {\cal F},_{x} .
\eqnum{9.9}$$
Since we assumed that rotation is  nonzero, we know that $g_{33} = - g_{\alpha \beta }w^{\alpha }w^{\beta }/n^{2} \neq  0$, and so we are allowed to  carry
out the coordinate transformation:
\medskip
$$z = z' - {\cal F}/g_{33} ,\eqnum {9.10}$$
\medskip
that, in virtue of (9.9), will lead to:
\par
$$
g_{13} = g_{23} = 0
\eqnum{9.11}$$
in the new coordinates. We have thus arrived at the metric form:
\par
$$
ds^{2} = (dt + y dx)^{2} - h(t,x,y) [dx + k(t,x,y) dy]^{2} - l(t,x,y) dy^{2} - A^{2}dz^{2},
\eqnum{9.12}$$
where $h, k$ and $l$ are functions to  be  found  from  the  Einstein
equations and {\it A} is an arbitrary constant.
\par
From now on, the allowed coordinate transformations are (5.6)
- (5.7), but with $T$ = const.
\par
The components of the Einstein tensor will now be referred to
the orthonormal tetrad implied by (9.12). The equation $G_{12} = 0$ is
integrated with the result:
\par
$$
k,_{t} = K(x,y)l^{1/2}/h^{3/2} - 1/\hbox{{\it h}.\qquad }
\eqnum{9.13}$$
The equation $G_{22} = \kappa p$, with $l$ elliminated by (9.6), is integrated
with the result:
\par
$$
h = [H^{2}(x,y) + K^{2}/(4\kappa p)]^{1/2} + H \sin  [2(\kappa p)^{1/2}t + \tau(x,y)],
\eqnum{9.14}$$
where $H(x,y), k(x,y)$ and $\tau(x,y)$ are arbitrary functions. Now $G_{11}
+ G_{22} = 2\kappa p$ imposes an additional condition on (9.13) and  (9.14)
that leads to $H = 0$ or:
\par
$$
H = (A^{2} - \kappa /p)^{1/2}K/(2\kappa ).
\eqnum{9.15}$$
The case $H = 0$ leads to the G\"odel solution  (see  below),  so  we
shall consider the more general case (9.15). Then, from (9.14):
\par
$$
h = [K/(2\kappa )] \{A + (A^{2} - \kappa /p)^{1/2} \sin  [2(\kappa p)^{1/2}t + \tau]\}.
\eqnum{9.16}$$
With such $h$, eq. (9.13) can be integrated with the result:
\par
$$
k = [2A(\kappa p)^{1/2}h]^{-1}(A^{2} - \kappa /p)^{1/2} \cos  [2(\kappa p)^{1/2}t + \tau] + L(x,y),\eqnum {9.17}
$$

\medskip

\noindent where $L$ is a new arbitrary function. The function $l$ is then calculated from (9.6), and an explicit solution of Einstein's equations is determined by (9.16) and (9.17).
\par
The transformations (5.6) - (5.7) with $T$ = const can  now  be
used  to  simplify  the  metric  tensor  so  that $L = 0.$   The
transformation that yields this is given in Appendix A. The still
allowed coordinate transformations that preserve the property $L =
0$ are given by (5.6)  -  (5.7)  with $T$  =  const  and  with  the
additional condition:
\par
$$
KF,_{x}F,_{y} + [\kappa /(A^{2}pK)] G,_{x}G,_{y} = 0.
\eqnum{9.18}$$
With $L = 0,$ from the equations $G_{01} = G_{02} = 0$ one obtains further:
\par
$$
\tau,_{x} = 2y(\kappa p)^{1/2} - A(p/\kappa )^{1/2}K,_{y} ,
$$
$$
\tau,_{y} = - (\kappa /p)^{1/2} K,_{x} /(AK^{2}).\eqnum {9.19}
$$
The integrability condition of (9.19) is:
\par
$$
(p/\kappa )^{1/2}(AK),_{yy} + (\kappa p)^{1/2}[1/(AK)],_{xx} - 2(\kappa p)^{1/2} = 0.
\eqnum{9.20}$$
By the same method as was used in Ref. 4 it can now be shown that
eq. (9.20) is at the same time the  integrability  condition  for
such a coordinate transformation (5.6) -  (5.7)  -  (9.18)  after
which (see Appendix A again):
\par
$$
K = (\kappa /A)y^{2},
\eqnum{9.21}$$
and then (9.19) implies:
\medskip
$$\tau = c = {\rm const}.\eqnum {9.22}$$

\medskip

\noindent The value of $c$ can be set arbitrarily by transformations of $t$ of the form $t = t'$ + const. To match Ref. 11 one should choose:
\par
$$
c = -\pi /2.
\eqnum{9.23}$$
Finally, the functions $h, k$ and $l$ in (9.12) are thus:
\par
$$
h = {1\over 2} y^{2}\{1 + [1 - \kappa /(pA^{2})]^{1/2} \cos [2(\kappa p)^{1/2}t],
$$
$$
k = {1\over 2} [(\kappa p)^{1/2}h]^{-1} [1 - \kappa /(pA^{2})]^{1/2} \sin [2(\kappa p)^{1/2}t],
$$
$$l = (4p^{2}A^{2}h)^{-1}. \eqnum{9.24}$$
\par
\medskip
\noindent This is equivalent under a simple  coordinate  transformation  to
the Ozsvath class III solution from Ref. 11.
\par
The velocity field $u^{\alpha } = \delta ^{\alpha }_{0}$ for  this  solution  has  nonzero
shear. The shear will vanish if and only if:
\par
$$
A^{2} = \kappa /p,
\eqnum{9.25}$$
and then the G\"odel solution in the form (5.11) results.
\par
The invariant definitions of the Ozsvath class III and of the
G\"odel solutions given at the  end  of  sec.  1  follow  from  the
derivation in this section.
\medskip

{\bf 10. Concluding remarks.}

These are the main results of the paper:

1. With nonzero rotation, any Killing field, existing for a metric whose matter source inherits the symmetry, must have the form (5.5) when represented in the Pleba\'{n}ski coordinates. When $\phi_{;\alpha} \neq 0$, the Pleba\'{n}ski coordinates can be adapted to the Kiling field so that $k^{\alpha} = {\delta^{\alpha}}_1$.

2. When two of the generators of the group are spanned on the velocity and rotation vector fields, while the third one is not, the collection of solutions with a dust source is exhausted by two sets:

a) The set of sec. 7, defined by a single differential equation (7.24), where the metric is (7.15) with $A = k_{13} = 0$, $F$ defined by (7.23) and $y(Y)$ defined by (7.21).

b) The set of sec. 8, where the metric is (7.15), with $b = 0$ and the metric functions are defined by an autonomous set of first-order equations (8.1) - (8.3), (8.8) and (8.10) (the integral in (8.8) and (8.10) can be calculated if the variable is changed as in (8.9)).

3. The solutions of Lanczos [1 and 19] and G\"{o}del [2] are limiting cases of both sets; their invariant definitions are given at the end of sec. 7 and of sec. 8.

4. With no symmetries pre-assumed, if the source is a rotating geodesic perfect fluid whose rotation vector field is covariantly constant, then the solution of the Einstein equations is the homogeneous (in four dimensions) Ozsvath class III solution [11]. If shear is zero in addition, then the G\"{o}del solution [2, see also 4] results.

Note the modification that the results 2a and 2b introduce in theorem 3.1 of King and Ellis [42]. Those authors considered spatially homogeneous models in which the velocity field of matter was tilted (i.e. was not orthogonal) with respect to the hypersurfaces of homogeneity. Theorem 3.1 says, among other things, that there are no tilted models of type I and that tilted models of type II have zero vorticity. Evidently, this does not apply to the case where the hypersurfaces of homogeneity are timelike. The solutions of sec. 7 are of Bianchi type II, they are "tilted" (because the velocity field is tangent to the hypersurfaces of homogeneity), yet rotation is not zero. The solutions of sec. 8 are tilted in the same sense, yet they are of Bianchi type I.

Other solutions that have been published earlier will be mentioned where appropriate in papers 2 and 3. A general overview of literature on related subjects will be included in paper 3.

\centerline {*****}

The algebraic calculations for this paper were done with the help of the program Ortocartan [43 - 44].
\par
\medskip
{\bf Appendix A. The transformation to $L = 0$ in (9.17).}
\par
A transformation of the class (5.6) - (5.7) with $T$  =  const
changes the functions $h$ and $k$ to such ones that can  be  cast  in
the form (9.16) and (9.17), respectively, with the new  functions
$K', \tau'$ and $L'$ expressed through the old ones as follows:
\par
$$
K' = K(F,_{x'} + LG,_{x'})^{2} + \kappa G,^{2}_{x'}/(A^{2}pK),\eqnum{A.1}
$$
$$
\tau' = \tau - 2(\kappa p)^{1/2}S + U, \eqnum{A.2}
$$
$$
L' = K'^{-1}[KF,_{x'}F,_{y'} + 2\hbox{{\it KLF}},_{y'}G,_{x'} + KL + KL^{2}G,_{x'}G,_{y'}
+ \kappa G,_{x'}G,_{y'}/(A^{2}pK)], \eqnum{A.3}
$$
where $S$ in (A.2) is the function from (5.6)  -  (5.7)  and $U$  is
determined by:
\par
$$
\cot  U = 2A(\kappa p)^{1/2}[2G,_{x'}(F,_{x'} + LG,_{x'})]^{-1}[- K(F,_{x'} + LG,_{x'})^{2}/(2\kappa )
+ G,^{2}_{x'}/(2A^{2}pK)]. \eqnum{A.4}
$$
Note that we are applying here (5.6) -  (5.7)  in  reverse,  i.e.
with the roles of $x^{\alpha }$ and $x'^{\alpha }$ interchanged. The functions  of  the
inverse transformation, denoted again by $S, F, G$  and $T$,  still
obey (5.7). For consistency of all the formulae it is  convenient
to choose $U$ from the segment $(\pi , 2\pi )$. Then,  the  limiting  cases
$G,_{x'} = 0$ and $F,_{x'} + LG,_{x'} = 0$ are included in (A.4) as the limits
$U = 2\pi $ and $U = \pi $ respectively (these limiting  cases  occur  when
$L,_{xx} = 0$ in the original coordinates).
\par
From (A.3), the equation $L' = 0$ turns out  to  be  consistent
with (5.6) - (5.7). In order to see this, one can solve (A.3) and
(5.7) for $F,_{x'}$  and $F,_{y'}$  and  then  impose  the  integrability
condition $F,_{\hbox{{\it x}'{\it y}'}} - F,_{\hbox{{\it y}'{\it x}'}} = 0.$ What comes out is  a  well-defined
(though highly nonlinear) partial differential equation of second
order for $G$ whose coefficients depend only on $K$ and $L$.
\par
The transformations preserving the property $L = 0$ are (5.6) -
(5.7) with (9.18), the latter easily follows from (A.3). Equation
(A.1) with $L = 0$  then  shows  how $K$  is  changed  by  such  a
transformation; this is useful in showing that coordinates  exist
in which $K,_{x} = 0$ (see the remark after (9.20)). The proof is 
identical as in Appendix C to Ref. 4. Note that the conclusion in
Ref. 4 is weaker than it could be: the $K(x)$ ($v(t)$ in  Ref. 4) is
determined up to an additive constant $C$. Hence, by a 
transformation of the form $y = y' +$ const and by  an  appropriate
choice of $C$ one can remove the linear and the constant terms in $K$ 
(resp. $v$) so that $K \propto  y^{2}$ (resp. $v \propto  t^{2}$ in Ref. 4).
\par
\medskip
\centerline{{\bf REFERENCES}
}
[1] K. Lanczos, {\it Z. Physik} {\bf 21}, 73 (1924).

[2] K. G\"odel, {\it Rev. Mod. Phys.} {\bf 21}, 447 (1949).

[3]  J.  Pleba\'nski,  {\it Lectures  on   non-linear   electrodynamics.} Nordita, Copenhagen 1970, pp. 107 - 115 and 130 - 141.

[4] A. Krasi\'nski, {\it Acta Phys. Polon.} {\bf B5}, 411 (1974).

[5] A. Krasi\'nski, {\it Acta Phys. Polon.} {\bf B6}, 223 (1975).

[6] A. Krasi\'nski, Acta Phys. Polon. {\bf B6}, 239 (1975) [available in extended form as a preprint].

[7] A. Krasi\'nski, {\it J. Math. Phys.} {\bf 16}, 125 (1975).

[8] A. Krasi\'nski, {\it Rep. Math. Phys.} {\bf 14}, 225 (1978).

[9] C. B. Collins, {\it Canad. J. Phys.} {\bf 64}, 191 (1986).

[10] H. Stephani, {\it Class. Q. Grav.} {\bf 4}, 125 (1987).

[11] I. Ozsvath, {\it J. Math. Phys.} {\bf 11}, 2871 (1970).

[12] S. Sternberg, {\it Lectures on  differential  geometry.} Prentice Hall, Englewood Cliffs, N. J. 1964, p. 141.

[13] G. F. R. Ellis, in: {\it General relativity and cosmology} (Proceedings of the International School of Physics "Enrico Fermi", Course 47). Edited by R. K. Sachs.  Academic  Press, New York and London 1971, p. 104.

[14] J. Ehlers, {\it Abhandl. Math. Naturw. Kl. Akad. Wiss. Lit. Mainz} {\bf 11}, 791 (1961); English translation in {\it Gen. Rel. Grav.} {\bf 25}, 1225 (1993).

[15] G. F. R. Ellis, {\it J. Math. Phys.} {\bf 8}, 1171 (1967).

[16] E. Verdaguer, {\it Phys. Reports} {\bf 229} no 1, 1 (1993).

[17] O. J. Bogoyavlenskii, {\it Methods in the qualitative theory of dynamical systems in astrophysics and gas dynamics.} Springer 1980.

[18] H. Stephani, {\it General relativity} (second edition).  Cambridge University Press 1990, pp. 284 - 288.

[19] K. Lanczos, {\it Gen. Rel. Grav.} {\bf 29}, (1997) (in press).

[20] J. P. Wright, {\it J. Math. Phys.} {\bf 6}, 103 (1965).

[21] W. J. van Stockum, {\it Proc. Roy. Soc. Edinburgh} {\bf 57}, 135 (1937).

[22] W. B. Bonnor, {\it J. Phys.} {\bf A13}, 2121 (1980).

[23] C. Hoenselaers, C. V. Vishveshwara, {\it Gen. Rel. Grav.} {\bf 10}, 43 (1979).

[24] A. H. Khater, M. F. Mourad, {\it Astrophys. Space Sci.} {\bf 163}, 247 (1990).

[25] S. K. Chakraborty, {\it Gen. Rel. Grav.} {\bf 12}, 925 (1980).

[26] M. Novello, M. J. Rebou\c{c}as, {\it Phys. Rev.} {\bf D19}, 2850 (1979).

[27] R. X. Saibatalov, {\it Gen. Rel. Grav.} {\bf 27}, 697 (1995).

[28] F. Bampi, C. Zordan, {\it Gen. Rel. Grav.} {\bf 9}, 393 (1978).

[29] J. N. Islam, {\it Proc. Roy. Soc. London} {\bf A353}, 523 (1977).

[30] J. N. Islam, {\it Proc. Roy. Soc. London} {\bf A385}, 189 (1983).

[31] P. Wils, N. van den Bergh, {\it Proc. Roy. Soc. London} {\bf A394}, 437 (1984).

[32] A. Georgiou, {\it Nuovo Cimento} {\bf B108}, 69 (1983).

[33] M. M. Som, A. K. Raychaudhuri, {\it Proc. Roy. Soc. London} {\bf A304}, 81 (1968).

[34] A. Banerjee, S. Banerji, {\it J. Phys.} {\bf A1}, 188 (1968).

[35] N. V. Mitskevi\v{c}, G. A. Tsalakou, {\it Class. Q. Grav.} {\bf 8}, 209 (1991).

[36] A. M. Upornikov, {\it Class. Q. Grav.} {\bf 11}, 2085 (1994).

[37] H. M. Raval, P. C. Vaidya, {\it Ann. Inst. Poincare} {\bf A4}, 21 (1966).

[38] M. J. Rebou\c{c}as, {\it Phys. Lett.} {\bf A70}, 161 (1979).

[39] M. J. Rebou\c{c}as, J. Tiomno, {\it Phys. Rev.} {\bf D28}, 1251 (1983).

[40] A. K. Raychaudhuri, S. N. Guha Thakurta, {\it Phys. Rev.} {\bf D22}, 802 (1980).

[41] A. Barnes, {\it Gen. Rel. Grav.} {\bf 4}, 105 (1973).

[42] A. R. King, G. F. R. Ellis, {\it Commun. Math. Phys.} {\bf 31}, 209 (1973).

[43] A. Krasi\'nski, {\it Gen. Rel. Grav.} {\bf 25}, 165 (1993).

[44] A. Krasi\'nski, M. Perkowski, {\it Gen. Rel. Grav.} {\bf 13}, 67 (1981).

\end{document}